# Intrinsic mixed Bloch-Néel character and chirality switch of skyrmions in asymmetric epitaxial trilayer

Pablo Olleros-Rodríguez†, Ruben Guerrero†, Julio Camarero†,‡, Oksana Chubykalo-Fesenko†,* & Paolo Perna†,*

† IMDEA Nanociencia, c/ Faraday 9, Campus de Cantoblanco, 28049 Madrid, Spain.

‡ Departamento de Física de la Materia Condensada, Instituto "Nicolás Cabrera" and Condensed Matter Physics Center (IFIMAC), Universidad Autónoma de Madrid, Campus de Cantoblanco, 28049 Madrid, Spain.

† Science Materials Institute (ICMM-CSIC), Campus de Cantoblanco, 28049, Madrid, Spain.

*Corresponding Authors: oksana@icmm.csic.es, paolo.perna@imdea.org

**Recent advances on the stabilization and manipulation of chiral magnetization configurations in systems consisting in alternating atomic layers of ferromagnetic and non-magnetic materials hold promise of innovation in spintronics technology. The low dimensionality of the systems promotes spin orbit driven interfacial effects like antisymmetric Dzyaloshinskii-Moriya interactions (DMI) and surface magnetic anisotropy, whose relative strengths may be tuned to achieve stable nanometer sized magnetic objects with fixed chirality. While in most of the cases this is obtained by engineering complex multilayers stacks in which interlayer dipolar fields become important, we consider here a simple epitaxial trilayer in which a ferromagnet, with variable thickness, is embedded between a heavy metal and graphene. The latter enhances the perpendicular magnetic anisotropy of the system, promotes a Rashba-type DMI, and can sustain very long spin diffusion length. We use a layer-resolved micromagnetic model (LRM) to describe the magnetization textures and their chirality. Our results demonstrate that for Co thickness larger than 3.6 nm, a skyrmion having an intrinsic mixed Bloch-Néel character with counter-clock-wise chirality is stabilized in the entire (single) Co layer. Noteworthy, for thicknesses larger than 5.4 nm, the skyrmion switches its chirality, from counter-clock-wise to clock-wise.**

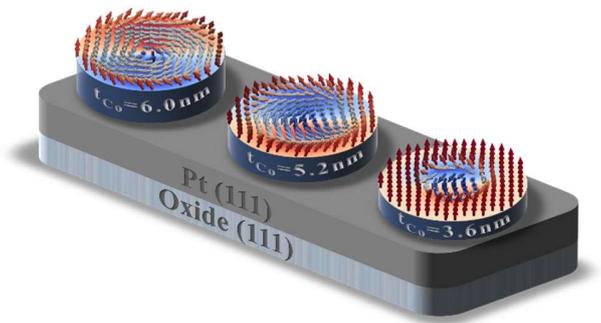

Spin-orbit interactions, along with low dimensionality and interfacial effects in ferromagnetic (FM) / heavy-metal (HM) systems have opened a possibility for a wide range of new applications in spin-orbitronics such as more efficient and lower power consuming (storage/processing) devices.[1] The benefits of these applications rely on the ability to nucleate, stabilize and manipulate whirling magnetization configuration, the so-called magnetic skyrmions.[1] Due to their small sizes (from a few to several hundred nanometers) and their intrinsic topological protection, these spin textures were proposed as candidates for their implementation in the next generation of memory and logic devices.[3] The chirality of these magnetic objects arises from the interplay between several energy contributions favoring either collinear or non-collinear spin configurations. The first contribution is related to the magnetic anisotropy energy (MAE). The second one is an antisymmetric exchange known as Dzyaloshinskii-Moriya Interaction (DMI) and is associated to the broken inversion symmetry at the interfaces between the FM and the HM layers.[3,4] Long-range dipolar interactions are also known to be of relevance in the stability, shape and sizes of the final skyrmionic states.[5,6,7] It was also reported that skyrmions can be moved by employing very low current densities, compared to the ones needed for domain walls (DWs) displacements[8] and that they can be electrically detected by means of anomalous Hall signals even at room temperature (RT).[10] The skyrmion motion can be driven either by spin-transfer[Error! Reference source not found.] or by spin orbit torque mechanisms,[Error! Reference source not found.,11] both relying on the spin orbit interaction (SOI).

From an experimental point of view the stabilization of the skyrmion spin textures can be achieved by engineering the materials, the interfaces composing the stacks and their thicknesses.[13] Nevertheless, real world spintronics applications require the engineering of materials in which chiral spin textures exist at (or above) RT and are protected from the atmosphere, while still amenable to reading and writing and, if possible, integrated with a material that can sustain very long spin diffusion length, such as graphene (Gr). Epitaxial Gr/Co/Pt(111) stack offers an ideal support for spin orbitronics technologies because of the high-quality interfaces obtained via the control of the thermal activated intercalation processes,[14] the enhanced perpendicular magnetic anisotropy, the possibility of effectively tuning the Dzyaloshinskii-Moriya interaction, and hosting the Néel-type domain walls (DWs) with counter-clock-wise chirality (CCW), controlled via the competing Rashba- and SOC-induced DMI at Gr/Co and Co/Pt interfaces respectively.[15]

The typical approaches employed to model macroscopically magnetic structures are based on micromagnetism that acts over the whole magnetization of the system and use effective (averaged) parameters.[16] It has been shown that depending on the complex interplay between parameters, skyrmions of either Néel or Bloch type are possible.[13,17] The magnetization confinement, for example in dots, creates an additional possibility to stabilize skyrmions due to the influence of magnetostatic interactions.[18] Importantly, very small skyrmions seem to be unstable against thermal fluctuations and in order to achieve thermal stability wide stacks of thin Co/Pt multilayers with many repetitions are commonly used.[19] The micromagnetic simulations have shown that the magnetostatic interactions can stabilize hybrid skyrmions which are of the Néel-type on the surface and of the Bloch-type in the stack center.[5] Here we report on hybrid skyrmions having the same mixed Bloch-Néel structure along the whole Co film thickness in asymmetric trilayers.

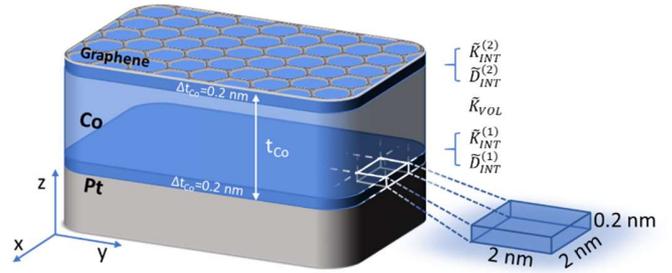

**Figure 1. Sketch of the Layer Resolved Model**. The magnetic volume is discretized into cells with the height of a monolayer of Co (=0.2 nm for FCC (111)-oriented structure). The micromagnetic parameters defined in each cell depend on their position along the z-axis, allowing to account for the interfacial nature of the perpendicular magnetic anisotropy and the Dzyaloshinskii-Moriya interaction.

The interfacial nature of SOIs, i.e., interfacial magnetic anisotropy, DMI, Rashba, .., should require complex atomistic calculations that generally imply large computation cost.[20] In this Letter, we implement a layer-resolved model (LRM) to account for the low dimensionality nature of the interactions that lead to macroscopic (averaged) parameters that depend on the thickness of the FM layer. The benefit of this model is that it allows us to correlate directly the modeling parameters to the experimental observables, which report macroscopic parameters. To validate the model, we reproduce the experimental characteristics of asymmetric Gr/Co($t_{Co}$)/Pt(111) epitaxial trilayer displaying large perpendicular magnetic anisotropy (PMA) and sizeable interfacial DMI,[15] as function of the Co thickness $t_{Co}$. Then, we study the properties of stable/metastable skyrmions in a nanodot geometry, evaluating the DW character (i.e. Bloch. or Néel-type) and the chirality of skyrmions varying $t_{Co}$.

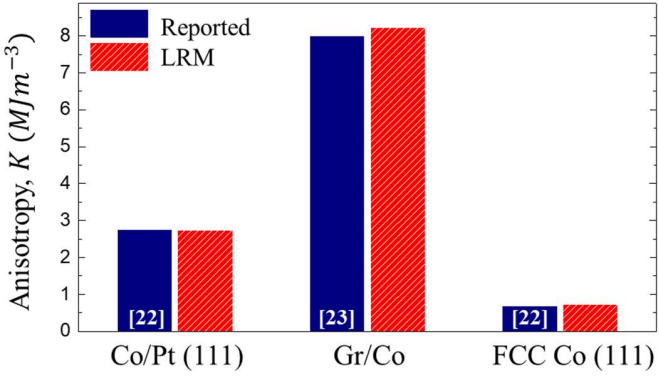

**Figure 2. Comparison of the anisotropy parameters obtained from the LRM model with those reported in the experiments.** The values of the magnetic anisotropy $K$ for Co/Pt(111) and FCC Co(111) are extracted from Ref.[22] and for Gr/Co from Ref.[23] The modeling parameters are obtained by assuming that the measurements of the effective MAE and DMI in Ref.[15,17] are the result of a weighted average of the parameters defined for each monolayer.

## The Layer-Resolved Model (LRM)

In order to model the trilayer Gr/Co/Pt(111) epitaxial systems, the magnetic volume of the FM layer (i.e., Co) is discretized into cells with dimensions 2 nm x 2 nm x 0.2 nm (length, width and height respectively). The dimension of the cell along the out-of-plane (OOP) direction is chosen to match with the thickness of a Co monolayer $\Delta t_{Co}$=0.2 nm.[15,21] The system is hence modelled atomistically along the OOP and micromagnetically along the in-plane (IP) directions, as schematically shown in **Figure 1**. This allows accounting for the interfacial SOI with no increased computational cost and for different parameters along z direction. Note that since the Co layers are strongly coupled via the exchange interactions, the continuum approximation for magnetization holds.

The Co cells at the top interface (i.e., Gr/Co) are described by the interfacial anisotropy constant $\widetilde{K}_{INT}^{(T)}$ and DMI $\widetilde{D}_{INT}^{(T)}$. The cells at the bottom interface (Co/Pt) are defined by $\widetilde{K}_{INT}^{(B)}$ and $\widetilde{D}_{INT}^{(B)}$ for the anisotropy constant and DMI respectively. The remaining Co volume, hereinafter referred as 'Bulk', is characterized by the anisotropy $\widetilde{K}_{VOL}$ of bulk FCC Co and by a null DMI. We can write the expressions for the weighted average of the magnetic anisotropy and the DMI of the systems respectively, as:

$$\langle \widetilde{K}_U \rangle = \frac{\left(\widetilde{K}_{INT}^{(T)}+\widetilde{K}_{INT}^{(B)}\right)\Delta t_{Co}+\widetilde{K}_{VOL}(t_{Co}-2\Delta t_{Co})}{t_{Co}} \quad (1)$$

$$\langle \widetilde{D} \rangle = \frac{\left(\widetilde{D}_{INT}^{(T)}+\widetilde{D}_{INT}^{(B)}\right)\Delta t_{Co}}{t_{Co}} \quad (2)$$

We assume that this weighted average of the layer-defined macroscopic parameters $\langle \widetilde{K}_U \rangle$ and $\langle \widetilde{D} \rangle$ are linked to the experimental observables (experimentally accessible quantities) $K_{eff}$ and $D_{eff}$ through $\langle \widetilde{K}_U \rangle - \frac{1}{2}\mu_0 M_S^2 = K_{eff}$ and $\langle \widetilde{D} \rangle = D_{eff}$.

In the case of an ideal symmetric Pt/Co/Pt trilayer $\widetilde{K}_{INT}^{(T)} = \widetilde{K}_{INT}^{(B)} = \widetilde{K}_{INT}^{Co/Pt}$, allowing to obtain the value of the uniaxial anisotropy constant of the Co/Pt interface ($\widetilde{K}_{INT}^{Co/Pt}$) from two independent measurements at different Co thickness. We refer to the experimental values of $K_{eff}$ reported in Ref.[15,17] for epitaxial trilayer of Pt/Co/Pt(111) and Gr/Co/Pt(111) with 0.6, 1.4 and 1.0 nm thick Co. We hence obtain $\widetilde{K}_{INT}^{Co/Pt} = 2.74 \times 10^6 \, J/m^3$, $\widetilde{K}_{VOL}^{Co} = 0.73 \times 10^6 \, J/m^3$ and $\widetilde{K}_{INT}^{Gr/Co} = 8.23 \times 10^6 \, J/m^3$. As it can be observed in **Figure 2**, by employing the aforementioned assumption, the values of the interfacial and volume anisotropies obtained in the LRM match almost perfectly with those resulting from experiments.[22,23]

For the determination of $\widetilde{D}_{INT}^{Co/Pt}$ and $\widetilde{D}_{INT}^{Gr/Co}$, we performed a similar approach. The effective DMI reported for the AlOx/Co(0.6nm)/Pt[15,17] is $D_{eff}^{AlOx/Co/Pt} = 1.47 \, mJ/m^2$ and by assuming $D_{INT}^{AlOx/Co} = 0$, we obtain $\widetilde{D}_{INT}^{Co/Pt} = 4.41 \, mJ/m^2$. Then, being the measured effective DMI of the

| SAMPLE | $t_{Co}$ (nm) | $K_{eff}$ (MJ/m³) | $D_{eff}$ (mJ/m²) |
|---|---|---|---|
| Pt/Co($t_{Co}$)/Pt [17] | 0.6 | 0.84 | 0.00 |
| Pt/Co(($t_{Co}$))/Pt [15,17] | 1.4 | 0.00 | 0.00 |
| AlOx/Co(($t_{Co}$))/Pt [17] | 0.6 | - | 1.47 |
| Gr/Co($t_{Co}$)/Pt [15] | 1.0 | 1.40 | 0.60 |
| **REGION** | | $\langle \widetilde{K} \rangle$ (MJ/m³) | $\langle \widetilde{D} \rangle$ (mJ/m²) |
| Pt/Co | | 2.743 | 4.41 |
| Gr/Co | | 8.232 | -1.41 |
| Bulk Cobalt | | 0.728 | 0.00 |

**Table 1. Values of the experimental ($K_{eff}$ and $D_{eff}$) and LRM calculated ($\langle \widetilde{K} \rangle$ and $\langle \widetilde{D} \rangle$) magnetic anisotropies and DMIs for different stacks.** Available experimental data (from Ref.[15,17]) and extracted micromagnetic parameters for different systems and Co thicknesses.

Gr/Co(1nm)/Pt $D_{eff}^{Gr/Co/Pt} = 0.6 \, mJ/m^2$ (from Ref.[15]) we

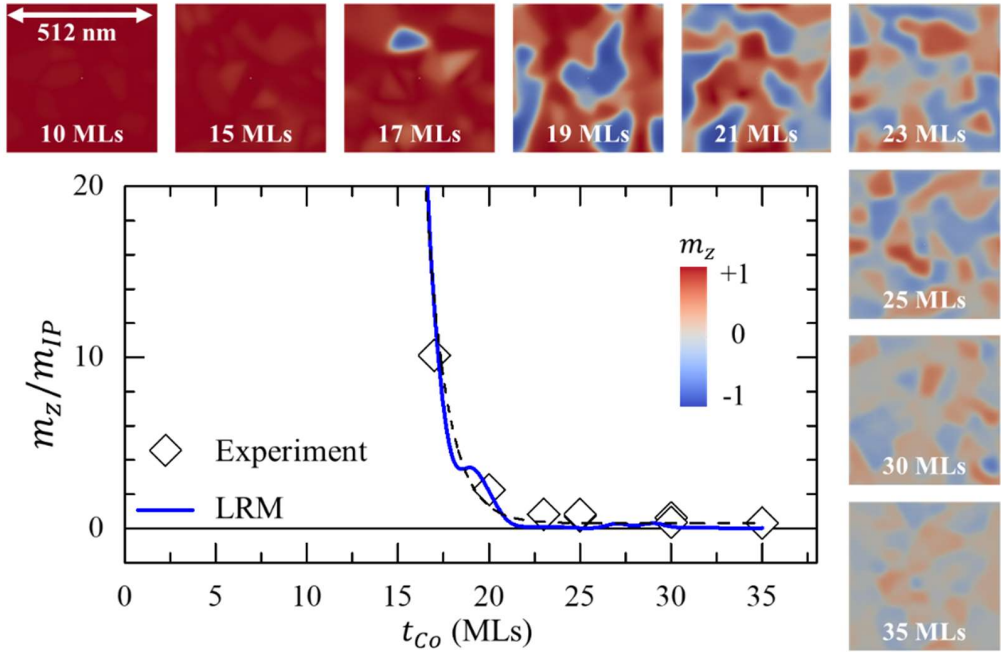

**Figure 3. Magnetization configurations and spin reorientation transition in Gr/Co($t_{Co}$)/Pt(111) simulated by LRM.** The main panel shows the ratio between the OOP and IP magnetization components at remanence, i.e. $m_z/m_{IP}$ as function of the Co thickness, $t_{Co}$. The diamond symbols refer to experimental data from Ref.[15], whereas the solid line presents the simulated ratio by LRM extracted from 512 x 512 nm² images of the magnetization configurations at selected $t_{Co}$ (in the insets).

can deduce the value for the DMI of the Gr/Co interface $\widetilde{D}_{INT}^{Gr/Co} = -1.41\ mJ/m^2$. **Table 1** lists the values of the experimental and estimated magnetic anisotropies and DMIs for different stacks.

It is worth noting that in our model the interface dependent energetic contributions are contained within the Co cells that are in contact with Pt and Gr. The DFT calculations show that these interactions, coming either from the inversion symmetry breaking at the interfaces or from the hybridization of the atomic/electronic orbitals, are extended to a few monolayers strongly decaying as a function of the distance from the interface.[24,25]

## Spin reorientation transition in Gr/Co($t_{Co}$)/Pt(111)

To test the validity of the proposed model, we have simulated the Co spin reorientation transition (SRT) as a function of the Co thickness in Gr/Co($t_{Co}$)/Pt(111) and compared it with the experiment.

First, we have calculated the ground state of 512 x 512 nm² samples with periodic boundary conditions (i.e. infinitely large thin films) with Co thicknesses ranging from 5 to 30 MLs. The simulations consisted in relaxing the system via energy minimization from the saturated state at 2 T to the remanent state at 0 T, and were carried out by the MuMax3 micromagnetic code.[26] The magnetization saturation was set as $M_{Sat} = 1.4 \times 10^6\ A/m$ (bulk value of Co). The exchange stiffness $A_{ex} = 24 \times 10^{-12}\ pJ/m$ and the effective damping parameters $\alpha = 0.3$ were taken from reported values.[27,28] To account for possible inhomogeneities in the surface, small variations of the uniaxial anisotropy vector were introduced in different regions of the surface.

**Figure 3** presents the results of the simulations. The images in the insets show the magnetization configurations at selected $t_{Co}$, from which we extract the ratio between the OOP and IP magnetization components at remanence, i.e. $m_z/m_{IP}$ as function of $t_{Co}$. This is shown in the main panel (blue line), together with the experimental values (diamonds symbols) from Ref.[15] The good agreement between the results of the simulations and the experimental values is clearly seen in the main panel of Figure 3. Particularly, the LRM model reproduces satisfactorily the smooth transition from OOP to IP states, as well as the critical thickness of Co for which the spin reorientation transition occurs. Note also that the magnetic configurations in the insets reproduce accurately the typical both macroscopic and nanoscopic images of PMA systems in the remanent state, i.e. labyrinth-like stripe domains.

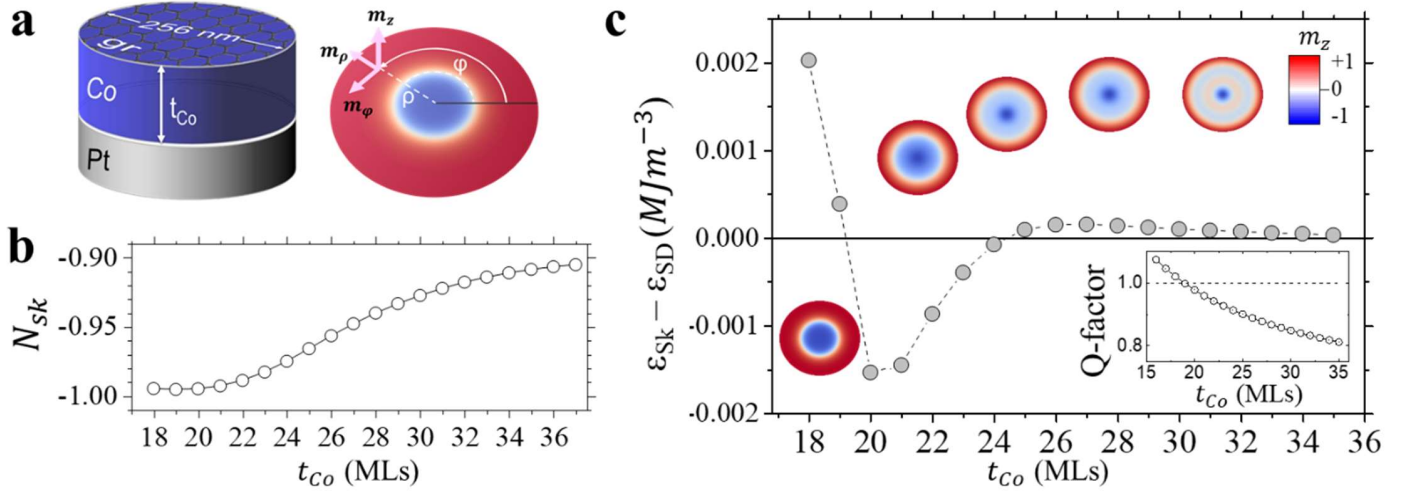

**Figure 4. Study of Skyrmion (Sk) in nanodots as function of the Co thicknesses.** Panel (a) presents the sketch of the Gr/Co/Pt nanodot geometry and the reference system. The Co thickness dependence of the skyrmion number ($N_{Sk}$) is shown in panel (b). Panel (c) displays the difference between the energy density of the system after the relaxation process starting from initial Sk ($\varepsilon_{Sk}$, blue symbols) and saturated single domain ($\varepsilon_{SD}$, red symbols) configuration. In the inset, the importance of the dipolar field in stabilizing the skyrmionic states is evidenced by the Q-factor vs. Co thickness plot.

## Study of the spin textures in nanodots

The nucleation and stabilization of magnetic skyrmions is improved by geometrical constraint. For instance, patterned arrays of magnetic nanodots have been proposed already as optimal architecture for skyrmion data storage and/or logic computation.[3,4,29-32]

We consider here Gr/Co($t_{Co}$)/Pt(111) nanodots with a diameter of 256 nm, as sketched in **Figure 4**(a). We have analyzed the magnetic configurations obtained starting from an initial state in which the magnetization at the core of the dot points antiparallel (along the OOP direction) with respect to the shell. The study has been made for Co thicknesses ranging from 5 to 35 MLs (i.e., from 1 and 7 nm). The system is then relaxed by means of an energy minimization process.

The results are resumed in panels b and c of **Figure 4**. For thicknesses below 18 MLs, we find that the core is absorbed and a saturated ground state (with magnetization pointing upwards along the OOP direction) is reached. For larger thicknesses, we obtain skyrmion (Sk) or pseudo-Sk states.

The difference between these two states relies on the structure of the spin texture surrounding the core, as discussed later. In both cases, we have computed the Sk-number ($N_{Sk} = \frac{1}{4\pi}\int \vec{m} \cdot \left(\frac{\partial \vec{m}}{\partial x} \times \frac{\partial \vec{m}}{\partial y}\right) dxdy$ ).[29] In panel b, we can see that from 18 to 22 MLs, $N_{Sk}$ takes values close to -1, while for larger Co thickness it is reduced to -0.9 due to the increase of the size of the Sk confined in the dot.

The Sk stability has been studied by computing the difference between the total energy density of the system after the relaxation process when starting from an initial Sk ($\varepsilon_{Sk}$, blue symbols in panel c) and a saturated single domain ($\varepsilon_{SD}$, red symbols) configuration. The Sk-state is hence stable for $20 \leq t_{Co} \leq 25$ MLs, otherwise it is metastable. The inset of panel c shows that for $t_{Co} > 18$ MLs the quality factor $Q = \langle\widetilde{K}\rangle(t_{Co})/K_{Dip} \leq 1$, where $K_{Dip} = \frac{1}{2}\mu_0 M_S^2$ is the shape anisotropy of an infinitely large thin film. These results evidence the importance of the dipolar field in stabilizing the skyrmionic states.

To get more insights into the properties of the spin textures in the dots, we have plotted the OOP component of the magnetization ($m_z$) along the diameter of nanodots, for Co thicknesses ranging from 18 to 37 MLs, i.e., for which the skyrmions are (meta)stable. The profiles are shown in panel b of **Figure 5**. This two-dimensional graphical representation evidences the magnetization changes along the radial distance within the dot. The plots in panel (a) are cut along the dot diameter at selected Co thickness, as indicated by the color bars in panel (b). At these specific thicknesses, the resulting spin configurations are shown in the left-side insets. From the graph, it results clear that only for $t_{Co} \leq 22$ MLs the profiles resemble the ones of a well-defined skyrmions known in case of constant DMI along the film thickness.[29-32]

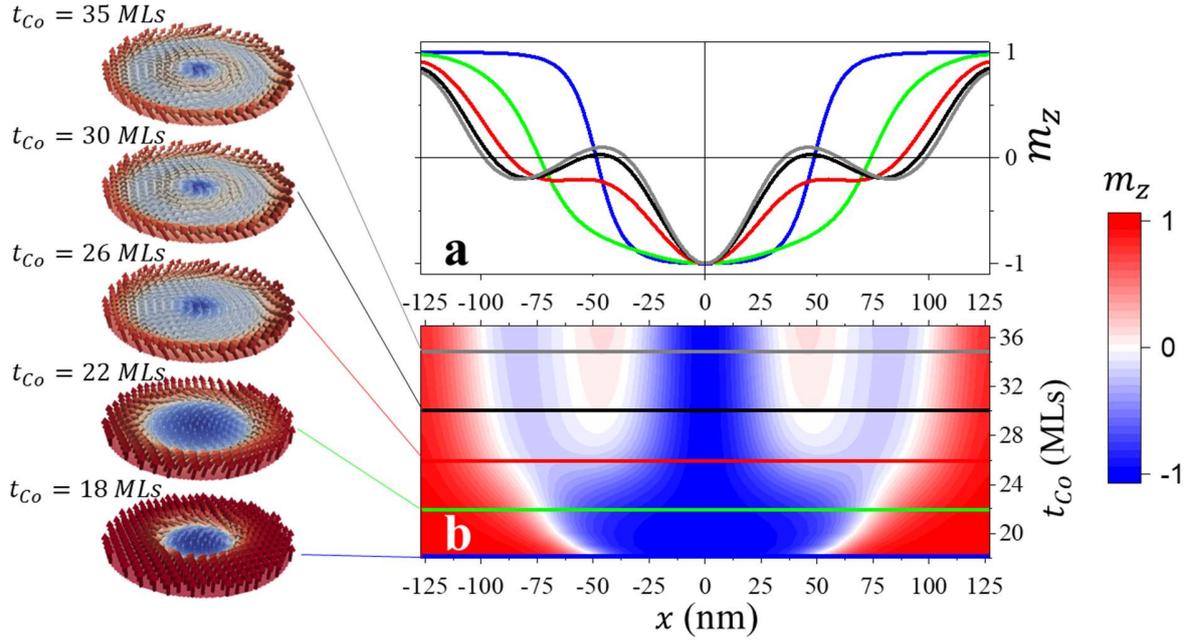

**Figure 5. Skyrmion profiles in nanodot as function of the Co thickness. (a)** pseudo-skyrmions profile ($m_z$) along the diameter (cross section along x direction) of the nanodots, and **(b)** two-dimensional map of the skyrmion and for Co film thicknesses from 18 to 37 MLs, for which the skyrmions are (meta)stable states. The plots in panel a are cut for selected Co thickness, as indicated by the color bars in panel b, whose spin configurations are shown in the left-side insets.

For larger thicknesses, i.e. $t_{Co} > 22$ MLs, the DW presents a complex structure. At 26 MLs (red curve in panel a), we even observe a change of the slope sign along the skyrmion. We refer to these states as pseudo-Sk due to the creation of two different pseudo DWs between the core and the shell of the dot. At 29 MLs (black curve) the spin wave-like behavior is more evident. We argue that this effect is due to the geometrical confinement imposed by the dot border and the core of the spin texture that compress the DW.

### Chirality of the mixed Bloch-Néel skyrmions

The chirality of the spin textures determines their dynamics under external excitations such as external fields or polarized electric currents.[33,34] In the following, we investigate the chirality of skyrmions in nanodots of Gr/Co($t_{Co}$)/Pt(111).

We consider a cylindrical coordinates reference system, as sketched in panel a of **Figure 4**. The OOP magnetization component is $m_z(\rho, \varphi, z)$, while the IP components are $m_\rho(\rho, \varphi, z)$ and $m_\varphi(\rho, \varphi, z)$. Due to the system high axial symmetry, the IP magnetization component $m_\rho$ ($m_\varphi$) is always normal (transversal) to the DW. With these notations, the radial component of the magnetization of a pure Bloch-type DW is equal to zero, i.e. $m_\rho(\rho, \varphi) = 0$. Conversely, pure Néel-type DW gives a vanishing transversal magnetization component $m_\varphi(\rho, \varphi) = 0$.

From the plots of the IP components of the magnetization along the diameter of the dot, in Figure 6, we can infer the character and the chirality of the DW (either Bloch- or Néel-type). The maps in panels (c) and (d) display $m_\varphi$ and $m_\rho$ as function of the radial coordinate and the Co thickness from which the specific magnetization variations and the skyrmions chirality are easily visualized. The plots in panel (a) and (b) are cuts along the radial direction at selected $t_{Co}$, as indicated by the color bars. Note that if the magnetization of the skyrmion core points along $-\hat{z}$ ($\hat{z}$), the non-zero $m_\rho(\rho, \varphi)$ and $m_\varphi(\rho, \varphi)$ are positive (negative) in case of counter-clockwise chirality and negative (positive) for the clockwise case.

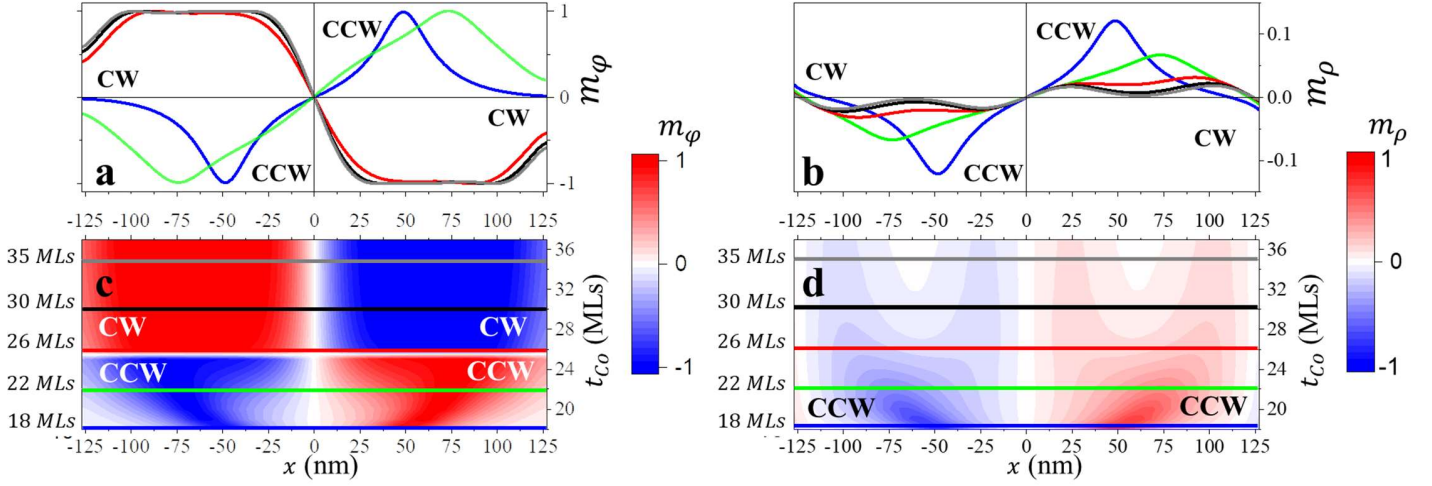

**Figure 6. Chirality of skyrmions in dots.** Cross section of the IP magnetization components $m_\varphi$ and $m_\rho$ (panels a,c and b,d respectively) as function of the Co thickness. The character and chirality of the DWs is evinced from the value and sign of the components. To note that $m_\varphi \gg m_\rho$ indicating the predominant Bloch contribution. The sign change of both $m_\varphi$ proves the chirality switch of the DWs for $t_{Co} > 26$ MLs, as discussed in the text.

In panels a and b, we can notice that: *i)* $m_\varphi$ and $m_\rho$ do not vanish completely, meaning that for all investigated Co thickness we obtain mixed magnetization configurations (i.e., mixed Bloch-Néel DWs); *ii)* $m_\varphi \gg m_\rho$ in the whole thickness range, indicating a predominant Bloch character of the skyrmions in Gr/Co($t_{Co}$)/Pt(111); and *iii)* for $t_{Co} >$ 22 MLs, the presence of multiple inflection points in $m_\varphi$ and $m_\rho$ indicate the creation of two pseudo-DWs within the dots, in agreement with the analyses of $m_z$ discussed in the previous section.

Noteworthy, the sign change of $m_\varphi$ directly proves the change of the chirality of the DWs for $t_{Co} > 26$ MLs. This behavior is due to the change of the slope sign of the $m_z$ profile in Figure 5, which is linked to the direction of the DMI field (see Ref.[26]). In particular, our results demonstrate that for $t_{Co} < 26$ MLs the Néel-type contribution is only <15% of the Bloch one. The analysis of the sign and intensity of $m_\rho$ and $m_\varphi$ indicates the presence

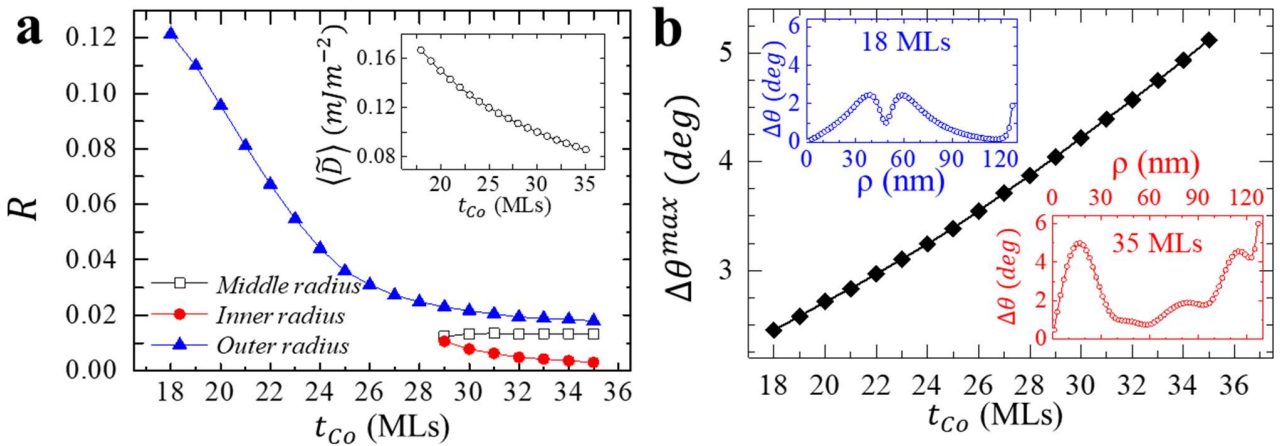

**Figure 7. Néel character contribution and angular deviation of the magnetization along the Co layer thickness.** Panel (a) shows the thickness dependence of R, defined for the point $m_z = 0$. For $t_{Co} \geq 29$ MLs, we define three radii corresponding to the point at which $m_z$ vanishes. The average DMI estimated from the model is shown in the inset. Panel (b) presents the maximum angular deviation ($\Delta\theta^{max}$) of the magnetization for each Co thickness with respect to the magnetization at the Gr/Co interface. These values are extracted from the radial angular deviation ($\Delta\theta$) as shown in the insets

of mixed DWs with CCW chirality having predominantly Bloch-type character. For larger $t_{Co}$ we observe the change of the sign of $m_\varphi$, indicating the transition of the chirality of the DW from Bloch-CCW to Bloch-CW.

To quantify the contribution of the Néel character of the DWs with respect to the Co layer thickness, we analyze the ratio:

$$R = \left.\frac{|m_\rho|}{\sqrt{m_\varphi^2 + m_\rho^2}}\right|_{m_z=0}$$

For $t_{Co} \geq 28$ MLs since the DWs have a complex structure, we define three radii, i.e. outer, middle and inner. As shown in panel a of **Figure 7**, R decreases with $t_{Co}$. This is an expected behavior as consequence of the reduction of the effective DMI (large DMI stabilizes Néel-type configuration) (see inset).

It is worth noting that the presented relaxed states have been obtained from an initial magnetization configuration without any defined chirality (see Supporting Information). Note also that mixed Bloch-Néel DWs have been recently reported in Co-based trilayers, and have been observed experimentally from bubble's expansion by Kerr microscopy.[35,36,37]

Finally, we have studied the magnetization dependence along the OOP direction of the whole Co layer within the nanodot. We have calculated the maximum angular deviation ($\Delta\theta^{max}$) of the magnetization at each Co thickness with respect to the magnetization at the top interface, i.e., Gr/Co. This is shown in panel b of Figure 7. We found that the largest $\Delta\theta^{max}$ is obtained at $t_{Co} = 35$ MLs, and is only ~5°. For all other thickness, the differences are below this value, indicating that the magnetization profile is practically homogeneous along the thickness. In the insets, we can observe that the deviation along the radial direction is not constant.

## Conclusions

In this work, we have demonstrated the importance of the surface character of the spin orbit interactions, like perpendicular magnetic anisotropy and Dzyaloshinskii-Moriya interaction, in determining the conditions for stabilizing chiral spin configurations. As a practical example, we have considered epitaxial Gr/Co($t_{Co}$)/Pt(111) trilayer, in which the large DMI at Co/Pt interface competes with the Rashba-type DMI at the interface with Gr.[15] This structure enables the realization of electrically controlled spin-orbitronic devices with high-quality interfaces and improved magneto-transport properties. The integration with Gr can sustain very long spin diffusion length.

We have implemented a layer-resolved model to account for the low dimensionality nature of the interactions, which lead to macroscopic parameters that depend on the thickness of the FM layer. We have demonstrated that our LRM model correctly reproduces the experimental magnetization configurations and the spin reorientation transition. In particular, we predicted the experimental parameters that will lead to Néel, Bloch or mixed chiral skyrmions. Our results demonstrate that for Co thickness larger than 3.6 nm intrinsic mixed (predominantly Bloch-type) skyrmions with CCW chirality are stabilized in 256 nm wide dots. For thickness larger than 5.4 nm, the skyrmions experience a drastic switch of chirality, from CCW to CW.

It is worth noting that in case of magnetic multilayers consisting in a large number of repetitions of $NM_1$/FM/$NM_2$,[5,38] the hybrid character reported is due to the different magnetization configurations in each FM layer. In this case, the dipolar interactions between the different layers give rise to a large total dipolar field that is ultimately responsible for the chiral spin arrangements together with the effective (constant) DMI and PMA along the FM thickness.

In our trilayer, instead, the *intrinsic* mixed (Bloch-Néel) character of the skyrmion exists in the entire (single) Co layer, whose macroscopic parameters have a realistic thickness dependence. Importantly, the chirality of the spin arrangements is easily switched from CCW to CW by simply varying the Co thickness.

## ACKNOWLEDGMENTS

The authors thank Prof. Kostyantyn Gusliyenko for fruitful discussion. This research was supported by the Regional Government of Madrid through Project P2018/NMT-4321 (NANOMAGCOST-CM) and by the Spanish Ministry of Economy and Competitiveness (MINECO) through Projects RTI2018-097895-B-C42 (FUN-SOC), FIS2016-78591-C3-1-R, FIS2016-78591-C3-3-R (SKYTRON). IMDEA Nanoscience is supported by the 'Severo Ochoa' Programme for Centres of Excellence in R&D, MINECO [grant number SEV-2016-0686]. The NVIDIA Corporation donated the GPU Quadro P6000 used for this research.

**Contributions**

P.O-R., O.C-F. and P.P. conceived and designed the project. P.O-R. developed the micromagnetic codes and performed the experiments. P.O-R., R.G., P.P. and O.C-F. analyzed the results. P.P., P.O-R. and O.C-F. prepared the manuscript with the help of J.C. All authors discussed and commented the manuscript.